\documentclass[prb,superscriptaddress,twocolumn,amsfonts,amsmath,floatfix]{revtex4-1}
\usepackage[sort&compress]{natbib}
\usepackage{color}      
\usepackage{epsfig}
\usepackage{dutchcal}
\usepackage[normalem]{ulem}
\newlength\figurewidth
\AtBeginDocument{\setlength\figurewidth{.5\linewidth}}

\def\kT{\ensuremath{k_\text{B}T}}
\def\Fext{\ensuremath{F_\text{ext}}}
\def\gammaeff{\ensuremath{\gamma_\text{eff}}}

\def\url#1{}

\begin{document}
\title{ Active and passive microrheology with large tracers in hard colloids }
\date{\today}
\def\unikn{\affiliation{%
  Fachbereich Physik, Universit\"at Konstanz,
  78457 Konstanz, Germany}}
\def\unial{\affiliation{%
  Departamento de Qu\'{\i}mica y F\'{\i}sica, Universidad de Almer\'\i{}a,
  04.120 Almer\'\i{}a, Spain}}
\def\unialinfo{\affiliation{%
  Departamento de Inform\'atica, Universidad de Almer\'\i{}a,
  04.120 Almer\'\i{}a, Spain}}
\author{F. Orts}\unialinfo
\author{M.~Maier}\unikn
\author{M.~Fuchs}\unikn
\author{G. Ortega}\unialinfo
\author{E.M. Garz\'on}\unialinfo
\author{A.~M.~Puertas}\unial

\begin{abstract}
The dynamics of a tracer particle in a bath of quasi-hard colloidal spheres is studied by Langevin dynamics simulations and mode coupling theory (MCT); the tracer radius is {varied from equal to} up to 7 times larger than the bath particles radius. In the simulations, two cases are considered: freely diffusing tracer (passive microrheology) and {tracer} pulled with a constant force (active microrheology). {Both} cases are connected by linear response theory for all tracer sizes. It links both the stationary and transient regimes of the pulled tracer (for low forces) with the equilibrium correlation functions; the velocity of the pulled tracer and its displacement  are obtained from the {velocity auto-correlation function} and the mean squared displacement, respectively. The MCT calculations give insight into the physical mechanisms: {At short times,} the tracer rattles in its cage of neighbours, with the frequency increasing linearly with the tracer radius asymptotically. The {long-time} tracer diffusion coefficient from passive microrheology, which agrees with the inverse friction coefficient from the active case, {arises from the transport of transverse momentum around the tracer. It} can be described with the Brinkman {equation for the transverse flow field obtained in extension of MCT, but cannot be recovered from the MCT kernel coupling to densities only}. The dynamics of the bath particles is also studied; for the unforced tracer the dynamics is unaffected, irrespective of the distance from the tracer. When the tracer is pulled, the velocity field in the bath decays with the distance from the tracer as $1/r^3$, as predicted by the Brinkman model, but different from the case of a Newtonian fluid.
\end{abstract}

\pacs{83.10.-y, 83.10.Rs, 64.70.pv}
\maketitle

\section{Introduction}
\label{sect_intro}

The response of a system to an external perturbation is one of the major open problems in statistical physics. In most systems, this response is complex and is divided in two regimes, the linear and the non-linear regime. In the former, the external perturbation is weak enough so that the effects on the system grow linearly with the external force, simplifying the theoretical analysis of the problem. To this category belong many well known approaches, such as the calculation of transport coefficients using Green-Kubo relations, or susceptibilities and Kramers-Kronigs relations \cite{Pottier2020}. The fluctuation-dissipation theorem, in its various forms, also derives from the linear response approximation. In general, this regime is well understood following the general theory of linear response and time correlation functions, as developed originally by R. Kubo \cite{Kubo1957}, and is well connected to statistical mechanics and thermodynamics \cite{Bianucci1995,Naze2020}. The non-linear regime, on the other hand, lacks a general formalism, and specific approaches have been developed for different systems.

Fundamental studies, either theoretical, experimental, or with simulations, have therefore focused on simple systems, to allow a deeper understanding or theoretical progress. A canonical example is the rheology of Brownian hard spheres, i.e. the mechanical (flow) properties of a fluid of hard colloids subject to a global tangential stress \cite{Larson1999,Bossis1989,Brady1993,Wagner2021}. For small stresses, the velocity (strain rate) field grows linearly with the external stress, allowing the definition of the system shear viscosity. For larger stresses, this linear regime crosses over to a non-linear shear thinning regime, with a sublinear growth of the strain rate with the stress (implying a decrease of the viscosity). Finally, for extremely large stresses, a shear thickening regime appears, with an increasing viscosity \cite{vanderWerff1989,Phung1996}. This complex behaviour arises from the interplay of the time and length scales induced by the external force and the intrinsic ones of the system in equilibrium (zero shear stress). 

Another simple problem, related to the previous one, is the dynamics of a soft matter system where a single particle, termed tracer, is pulled by an external force \cite{Cicuta2007,Wilson2011,Furst2017}. The effective friction can be obtained from the steady tracer velocity at long times, and a linear regime is obtained for small forces, where the tracer velocity is proportional to the force. For larger forces, a non-linear, force-thinning, regime is observed. This case is known as active microrheology, and experimentally this is achieved by inserting tracers which respond to an external field in an inert bath or host system, typically of soft matter, such as colloidal systems \cite{Habdas2004,Wilson2009,Sriram2010}, but also living cells \cite{Wirtz2009,Nishizawa2017,Wu2020}, or soft food \cite{Lu2016,Yang2017}, or metallic glasses \cite{Yu2020}. The theory approaches are more involved than in rheology because the strain field in the host system is non-affine \cite{Squires2008,Puertas2014}. On the other hand, the zero-force case is known as passive microrheology and concerns the diffusion of the tracer particle freely in the complex host medium \cite{Mason1995,Guzman2018}. The connection between active and passive microrheology relies on linear response \cite{DePuit2011,Leitmann2018,Caspers2023}. For low densities, non-linear active microrheology is studied with an appropriate description of the microsctructural deformation and stress imbalance in the bath \cite{Squires2005,Carpen2005,Khair2006}; the connection is then established as a small-deformation limit. For dense suspensions, the Smoluchowski equation and mode coupling theory has been used to describe the dynamics \cite{Puertas2014,Gazuz2009,Gazuz2013}, although the connection with bulk hydrodynamics was discussed only recently.

In this paper, we consider simulations of passive and active microrheology in a dense bath of hard colloids. The tracer particle is also hard, and its size is varied from similar to the bath particles to seven times larger. The equilibrium, linear and non-linear regimes are studied, checking the validity of the linear response approximation and the effects on the bath. Additionally,  calculations within generalized hydrodynamics and mode coupling theory are performed in order to identify the underlying physical mechanisms. A previous paper on this system \cite{Orts2020} focused on the linear regime and compared the results with the Brinkman fluid \cite{Brinkman1947}, although our interpretation of the terms in the equation is different from the original model. The linear response approximation was tested with the relation between the diffusion coefficient in the transversal direction, and the friction coefficient from the steady velocity--force relation. Here, this analysis is extended to passive microrheology (zero force) and the non-linear regime (high forces). The tracer mean squared displacements and VACFs in equilibrium (passive microrheology) are used to calculate the response of the system in the linear regime, and the predictions are tested. The non-linear regime is noted by deviations from this prediction. Finally, the effects of the moving tracer on the bath, are studied.

\section{Simulation details}
\label{sect_sim}

The system is composed of one tracer of radius $a_t$ and $N-1$ polydisperse bath particles of average radius $a$ -- the radii of the bath particles follow a flat distribution of width $0.1\,a$. All particles, including the tracer, have the same mass, $m$, and friction coefficient with the solvent, $\gamma_0$, and undergo Langevin microscopic dynamics, i.e. the equation of motion for particle $j$ reads \cite{Dhont1996}:

\begin{equation}
m \frac{d^2\, {\bf r}_j}{dt^2}\:=\: \sum_{i\neq j} {\bf F}_{ij} - \gamma_0 \frac{d\, {\bf r}_j}{dt} + {\bf f}_j(t) + {\bf F}_{ext} \delta_{j1}\;,
\label{Langevin}
\end{equation}

\noindent where ${\bf F}_{ij}$ is the interaction force between particles $i$ and $j$, ${\bf f}_j$ is the random Brownian force, which fulfills the fluctuation-dissipation theorem, $\langle {\bf f}_j(t) \cdot {\bf f}_j(t') \rangle = 6 k_BT \gamma_0 \delta(t-t')$, where $\kT$ is the thermal energy and $\delta(x)$ is the Dirac-delta symbol \cite{Dhont1996}, and the external force, ${\bf F}_{ext}$, acts only on the tracer (as shown by the Kronecker-delta symbol, $\delta_{j1}$). 

The interaction force ${\bf F}_{ij}(r)$ derives from the inverse-power potential:

\begin{equation}
V_{ij}({\bf r})\:=\:\kT \left( \frac{r}{a_{ij}} \right)^{-36}\;, \label{potential}
\end{equation}

\noindent which mimics the hard-core repulsion \cite{Lange2009}. {Note that we continue to speak about hard-core interactions in the following.} Here $r=\left| {\bf r} \right|$ is the center to center distance between the particles, and $a_{ij}=a_i+a_j$. Periodic boundary conditions are used in all cases, and the center of mass of the system is not fixed. 

In our simulations, the mean bath particle radius, $a$, is the unit of length, the thermal energy $k_BT$, the unit of energy and $m$ is the unit of mass. The friction coefficient with the solvent is set to $\gamma_0=5\,\sqrt{m \kT}/a$, and the Langevin equations of motion are integrated using the Heun algorithm \cite{Paul1995}, with a time step of $\delta t=0.0005 \, a \sqrt{m/\kT}$. Note that a microscopic time scale is defined by $a \sqrt{m/\kT}$, while the diffusive time scale is given by $a^2/D_0 = a^2 \gamma_0/kT$. Hydrodynamic interactions have been neglected for simplicity; although they are known to modify the results, these effects are more important at lower densities, and only quantitative changes are detected \cite{Nazockdast2016,Marenne2017,Chu2019}.

In passive microrheology, ${\bf F}_{ext}=0$, whereas ${\bf F}_{ext}>0$ in the simulations of active microrheology. In the latter case, the system (with the tracer) is equilibrated without external force and for $t\geq 0$ the contant force pulls the tracer. The effective friction coefficient, $\gammaeff$, is obtained from the steady state relation $\Fext = \gammaeff \langle v \rangle$, where $\langle v \rangle$ stands for the average tracer velocity. In contrast, in passive microrheology, the diffusion coefficient is calculated from long time mean squared tracer displacement (MSD), or the integral of the velocity autocorrelation function (VACF). 

In order to study finite size effects in the simulations, different systems, with different numbers of bath particles have been considered: $N=216$, $512$, $1000$, $2197$, $4096$, $8000$, $15625$ and $32768$ and the diffusion coefficient or friction coefficient is studied. The volume fraction of the bath is $50\%$ irrespective of the number of particles, and the tracer volume is not considered, following our previous analysis \cite{Orts2019}. Because this procedure needs to be repeated for more than $500$ independent trajectories for all system sizes, external forces and tracer sizes, high performance computing has been used. Two codes were prepared, one in FORTRAN, to be run in CPU-cores, and another one in CUDA, which is executed in GPU-cores \cite{Ortega2017}. The distribution of tasks is optimized with a genetic algorithm, ensuring that all CPU- and GPU-cores in our computing cluster finish their assigned number of tasks almost simultaneously \cite{Orts2020-JSC}. 

The simulation procedure requires a pre-processing step where the runtime of a single trajectory for every case and core must be determined. The genetic algorithm then provides the optimal distribution of tasks for every core, which is actually run in the cluster. The results for the diffusion coefficient (in passive microrheology) or friction coefficient (in active microrheology) are plotted as a function of the inverse system size, $1/L$, following the prediction from hydrodynamics \cite{Hasimoto1959}. As shown below, the finite size effects do not follow this trend, and even become independent of the system size for large enough systems. Therefore, further simulations to analyze in detail the properties of the tracer dynamics are run using systems beyond this critical size where the dependence on the number of bath particles has disappeared.

Further details on the system or simulation procedure are given elsewhere \cite{Orts2019,Orts2020}. 

\section{Mode coupling theory}
\label{sect_mct}

Mode coupling theory (MCT) has been used to rationalize the results of passive microrheology. This allows to demonstrate whether structural rearrangements on the length scale of the average bath particle separation, the so-called {\it cage effect}, determine the tracer motion. It is well established that the dynamical correlations resulting from the cage effect  are semi-quantitatively captured by MCT for not too high packing fractions \cite{Goetze,BinderKob}.

\subsection{Tracer dynamics}

In order to gain insight into the microscopic mechanisms determining the tracer dynamics, a binary mixture of hard spheres is considered within MCT.
The relative packing fraction of the large particles is chosen to be extremely small, so that the dilute limit of tracer particles is obtained. 
The resulting equation for the VACF of a tracer particle,  $K(t) = \langle {\bf v}(t) \cdot {\bf v} \rangle /(3v^2_{th})$ with thermal velocity $v^2_{th}=k_BT/m$, then reads \cite{Mandal2019}:

\begin{equation} 
\frac{dK(t)}{dt}+\Gamma K(t)+v_{th}^2\int_0^tdt'~\mathcal{m}(t-t')K(t')=0 \;. \label{MCT-K}
\end{equation}
Here, the rate $\Gamma$ models the damping from the background solvent included in the Langevin dynamics, and the memory kernel $\mathcal{m}(t)$ encodes the retarded friction on the tracer arising from the surrounding bath particles.  This is a nontrivial function determining the tracer motion which is approximated by MCT as functional of the bath $\phi_q(t)$ and tracer $\phi^t_q(t)$ density correlators \cite{Goetze}: 
\begin{equation} \label{eq1b}
\mathcal{m}(t)=\int_{0}^{\infty}dq~v(q)\; \phi_q(t)\; \phi_q^t(t) \;,
\end{equation}
with the so-called vertex
\begin{equation}
v(q)=\frac{1}{6\pi^2 \, n}\frac{(q^2S_q^{tb})^2}{S_q}\;,
\end{equation}
\noindent with $n$ the bath particle density, $S_q$ the bath structure factor and $S_q^{tb}$ the equilibrium tracer-bath density correlations, both depending on  the wavenumber $q$. The non-Markovian friction in $\mathcal{m}(t)$ arises from the bath density fluctuations close to the tracer. They are decomposed into wavevector modes and then become the product of a bath density times a tracer density fluctuation. In MCT, the correlation function of these pair-fluctuations is factorized into the product of bath $\phi_q(t)$ and tracer $\phi^t_q(t)$ density correlators as would be valid for Gaussian variables. Both functions are normalized to unity at $t=0$. The dynamics of both correlators is then determined from the self-consistent MCT equations of motion that are derived with similar factorization approximations \cite{Goetze}. The vertex $v(q)$ gives the force strength of a density fluctuation with wavevector $q$. Note that while the bath structure factor $S_q$ is independent on the tracer properties, $S_q^{tb}$ changes strongly with tracer size, as shown below (Fig. \ref{sq}) in comparison with the simulation results. 
In summary, MCT derives the dynamics from the equilibrium structure, which shall here be taken from the well-established Percus-Yevick (PY) approximation for the static structure factor \cite{Baxter1970}. We checked that the relative packing fraction concentration of larger spheres, $\hat{x}_{large}=0.002$, gives the dilute limit. Note that a monodisperse hard sphere bath is analysed theoretically, while a polydisperse one is simulated; for a discussion of differences see \cite{Weysser}.
The numerical MCT calculations are then completely specifed by stating the discretization. An equidistant wavevector grid is used with $q_i=(i-0.5)\Delta q$ for $\Delta q = 0.2/a$ and $i=1,\ldots,200$. For details on the theory and the numerical procedure, see Ref.~\cite{Voigtmann2003}.

\subsubsection{Asymptotic limit of large tracer sizes}

The passive motion of the tracer changes characteristically with its size because it interacts with the bath particles across its surface area, which increases with $a_t^2$.  This can be recognized from the MCT kernel $\mathcal{m}(t)$ when entering the asymptotic limit of the tracer-bath partial structure factor \cite{Kroy2002}. This is the Fourier-transform of the tracer-bath particle total correlation function $h^{tb}({\bf r})$  in real space, which becomes a step-function for large tracers, $h^{tb}({\bf r})\to - \Theta(\sigma-r)$ for $a_t \to \infty$, with the Heaviside step function $\Theta(x)$ and distance at contact $\sigma=a_t+a$. Particle overlaps are prohibited, and the bath particle density changes from zero inside the tracer to the average  value outside it \cite{Baxter1970,Hansen1986}. Fourier-transforming, the partial structure factor follows in this limit:
 $S^{tb}_q = -4\pi n\ \sigma^3\, f(q\sigma)$,
where $f (x) = j_1(x)/x =(\sin(x) - x \cos(x))/x^3$, with $j_1(x)$ the first spherical Bessel function. 
This enables one to determine the large tracer size limit of the MCT memory kernel:
\begin{equation}\label{eq2}
\mathcal{m}(t) \to  \frac{8n\sigma^6 }{3} \frac{\phi_0(t)}{S_0}\;  \int_{0}^{q_c} dq\, q^4 \;f^2(q\sigma) \; \phi^t_{q}(t)\;.
\end{equation}
Here, it was used that the bath functions do not vary with $q\sigma$ and can  effectively be evaluated at $q=0$. Also, a cut-off $q_c$ was introduced anticipating that the integral in Eq.~\eqref{eq2} may not converge after the neglect of the bath correlations. 

\subsubsection{Short time vibrational phenomena}

Taking a time-derivative of the equation of motion \eqref{MCT-K} of the VACF, $K(t)$, one recognizes the equation of a damped harmonic oscillator with vibration frequency $\omega$. It is given by $\omega^2 = v^2_{th} \mathcal{m}(0)$  assuming that the memory kernel varies slowly compared to $K(t)$, and can be estimated  from Eq.~\eqref{eq2} as
\begin{equation}
\omega^2= 
\frac{\sigma^2 v^2_{th}  }{a^4}\; \big[  \frac{\varphi}{\pi} \frac{q_ca}{S_0} \big]\quad\mbox{for }\; a_t\to\infty \; .  \label{MCT-w}
\end{equation}
Here, the asymptote  $\int_{0}^{q_c\sigma} dx~ x^4 f^2(x)\approx \frac 12 q_c\sigma $ was used, and the packing fraction $\varphi=\frac{4\pi}{3} n a^3$ was introduced. The force from the bath particles pushing back  the tracer scales with the surface area of the tracer. This large-tracer limit will be compared to the Langevin simulations in the passive case below.

For the numerical study of the short-time vibrational properties, the Newtonian limit is assumed in the equations of motion of the correlators $\phi_q(t)$ and $\phi_q^t(t)$; {this neglect of the Langevin friction in the bath particle motion is done as purely technical simplification and, we expect,  does not affect the investigation of the vibrational frequency. The expectation is based on the aspect that the vibrations  start already at short times during the ballistic motion of the tracer.} Also for simplicity, the thermal velocities are set equal, viz.~$v_{th}^{a}=v_{th}=v_{th}^{a_t}$. Comparing the numeric MCT solutions to the asymptotic law~\eqref{MCT-w} gives the unknown cut-off $q_c=7.5/a$. In order to capture the Langevin damping of the tracer vibrations in the simulations, $\Gamma$ in Eq.~\eqref{MCT-K} is adjusted when comparing to the VACFs in Fig.~\ref{vcorr} below. Note, that the characteristic time set by the bath remains $a/v_{\rm th}$, however.

\subsubsection{Long time diffusional phenomena}

For comparisons with the diffusion processes in the Langevin system, the Brownian limit of the MCT equations is taken. This corresponds to $(i)$ neglecting the inertial term in Eq.~\eqref{MCT-K}, viz.~setting $dK(t)/dt\approx0$. The ratio $v^2_{th}/\Gamma=D_0^{a_t}$ then determines a short time diffusion coefficient of the tracer. Also $(ii)$, it corresponds to using Brownian equations of motion for the bath particles so that the  characteristic time  becomes $a^2/D_0^{a}$. For simplicity, 
the short time diffusion coefficients are taken identical, viz.~$D_0^{a}=D_0=D_0^{a_t}$.

The most interesting dynamical tracer function then becomes the MSD, $\langle \delta r^2(t)\rangle$ which is connected to the VACF via  $ \partial_t^2 \langle \delta r^2(t)\rangle = 2 K(t)$ and obeys the MCT equation of motion
\begin{equation}\label{eq4}
    \langle \delta r^2(t)\rangle + D_0 \int_0^tdt'~\mathcal{m}(t-t')\;  \langle \delta r^2(t')\rangle = 6 D_0 \, t \; ,
\end{equation}
\noindent where the memory kernel of Eq.~\eqref{eq1b} appears again.

For long times, the tracer diffuses. This identifies the effective diffusion constant $D^t=\lim_{t\to\infty} \langle \delta r^2(t)\rangle / (6 t)$. Equation~\eqref{eq4} leads to \cite{Mayr1998}
\begin{equation}\label{eq5}
D^t = \frac{D_0}{1+D_0 \int_0^\infty dt\;  \mathcal{m}(t)}\;.     
\end{equation}
For large tracer sizes, the tracer correlator in the integral of Eq.~\eqref{eq2} is slow compared to the bath one. Denoting the bath structural relaxation time as $\tau=\int_0^\infty dt\phi_0(t)$, the asymptotic limit of $D^t$  is found via Eqs.~\eqref{eq2} and \eqref{eq5}. This predicts $D_t\propto a^4/(\sigma^2 \tau)$ for $a^t\to\infty$. This MCT prediction is based on the approximation that the bath density fluctuations determine the long-time tracer diffusion. This is appropriate at high density close to glassy arrest. Yet, as is well known from the long time tails discovered for a single colloid in a Newtonian solvent \cite{Hansen1986}, shear current fluctuations dominate the viscous forces in a regular fluid state. Thus the prediction for $D^t$ in Eq.~\eqref{eq5} should not be expected to hold at the rather low densities studied in the simulations, where transversal flow properties of the bath particle fluid need to be considered. They are coupled to stress fluctuations, which recently were studied in Langevin systems using approaches linked to MCT \cite{Maier2017,Maier2018,Vogel2020}. This approach shall be summaried in the next section and leads to the so-called Brinkman equation for the transverse degrees of freedom of a Langevin fluid.

\subsection{Friction force correlations}

The Zwanzig-Mori calculations that are the starting point of MCT, were recently generalized to stress fluctuations in viscoelastic fluids \cite{Maier2017,Maier2018} and the generalized hydrodynamic equations were derived for Langevin fluids \cite{Vogel2020}. The complete spatio-temporal structure of stress correlations,  including shear ones, were treated, while MCT considered the transport kernels of longitudinal and transverse fluctuating stresses only. 
Assuming  incompressible flows, the derived equations of motion can be coarse-grained in the hydrodynamic limit to the Brinkman equation \cite{Brinkman1947,Vogel2019} that will be used below for analyzing the tracer friction.  This hydrodynamic equation (given in Eq.~\eqref{brinkman} below) generalizes the Navier-Stokes equation by including a friction term that the tracer experiences from the Langevin background solvent. If this term is small, Stokes friction holds where $D^t \propto \sigma^{-1}$ because of the long-ranged shear velocity field of the bath particles around the tracer. If the Langevin friction dominates, the volume of the tracer experiences a drag force from the Langevin damping, which gives $D^t \propto \sigma^{-3}$. Both limits are contained in the hydrodynamics of Ref.~\cite{Vogel2019}, where a cross-over length $\xi=\sqrt{\eta/(n\gamma_0)}$ appears that grows like the square root of the viscosity with increasing density.

Inspired by MCT, the study of stresses has also been generalized with non-Markovian kernels so that viscoelasticity is captured and the limit of a deforming disordered solid can be discussed. This requires MCT approximations for the (longitudinal and shear) viscosity kernels  \cite{Goetze} instead of for the tracer friction memory kernel $\mathcal{m}(t)$.
We will refrain from these additional MCT calculations and below compare the fitted $\xi$ to the one taking $\eta$ from the simulations.


\section{Results and discussions}
\label{results}

The results from passive microrheology are presented in the first place with tracers of different sizes, and compared with the theory model of Section III.
The case of forced tracers is presented next. For low forces, the comparison of both cases tests the validity of linear response. 

\subsection{Passive microrheology}

\subsubsection{Static structure}

\begin{figure}
\psfig{file=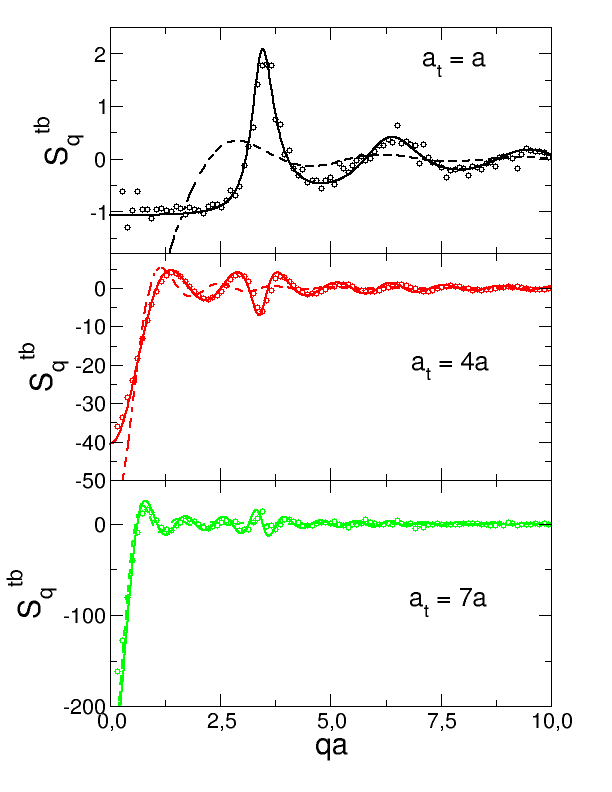,width=0.95\figurewidth}
\caption{Comparison between PY theory (lines) and simulations (symbols)  of the tracer-bath structure factor for different tracer sizes, for different panels. Note that in the simulations, the bath volume fraction is $\phi=0.50$, while in the theory $\phi=0.48$. The dashed lines correspond to the approximation $S^{tb}_q = - 4\pi n \sigma^3 f(q\sigma) = - 3 \phi (\sigma/a)^3 f(q\sigma)$ in every case.}\label{sq}
\end{figure}

The free energy of inserting the tracer is related to the tracer-bath structure factor \cite{Hansen1986}. 
It encodes the packing of the bath particles around the tracer. Moreover, the partial structure factors are the only inputs into the MCT dynamics, and in particular, the tracer-bath one is the most interesting,  as it varies strongly with tracer size. In the theory it is calculated using the Percus-Yevick (PY) approximation for mixtures, with an extremely low density for the large spheres. Fig. \ref{sq} shows a comparison of the calculated partial structure factor and the results from simulations. In order to improve the comparison, the density of the system is adjusted in the theory, a common issue in the theory, that partly explains the difference between the predicted glass transition within Mode Coupling Theory and experimental (or simulation) hard spheres. Here, the optimal comparison is found for a volume fraction $\phi=0.48$, where the PY approximation describes correctly all the salient features of the tracer-bath structure for all the tracer sizes (recall that the volume fraction in the simulations is $\phi=0.50$). The small $q$-limit of the partial structure factor shown in Fig.~\ref{sq} can be interpreted as a $q$-dependent chemical potential or solvation energy  for inserting the tracer. Its scaling 
$S^{tb}_q \propto - n\sigma^3 f(q\sigma)$ for $a_t\gg a$, where $f(x)$ has a deep minimum at $x=0$ can be used to explain the size dependence of the tracer motion in the following.

The dashed lines in the figure show the approximation $S^{tb}_q = - 4\pi n \sigma^3 f(q\sigma)$, used in the calculation of the memory kernel, Eq. (\ref{eq2}). Note that the MCT friction kernel \eqref{eq1b} asymptotically gets dominated by wavevectors below the peak in the bath structure factor, where the approximation improves for large tracer size. For large wavevectors, on the other hand, the approximation fails (see in particular the dip at $qa \approx \pi$) but the contribution of this range of $q$ is not dominant.
 
\subsubsection{Finite size analysis}

We move now to the analysis of the dynamical properties of the tracer particle. Long range hydrodynamic interactions cause finite size effects (FSE) in simulations of a single particle moving in a viscous Newtonian fluid due to the interaction with its periodic images, as shown theoretically by Hasimoto \cite{Hasimoto1959}. The inverse friction coefficient, or diffusion coefficient, depends linearly on the inverse simulation box length, and the slope is set by the viscosity of the fluid. Simulations of passive microrheology in a bath of hard spheres with microscopic Newtonian dynamics have tested this prediction with good agreement \cite{Yeh2004,Sokolovskii2006}. 

\begin{figure}
\psfig{file=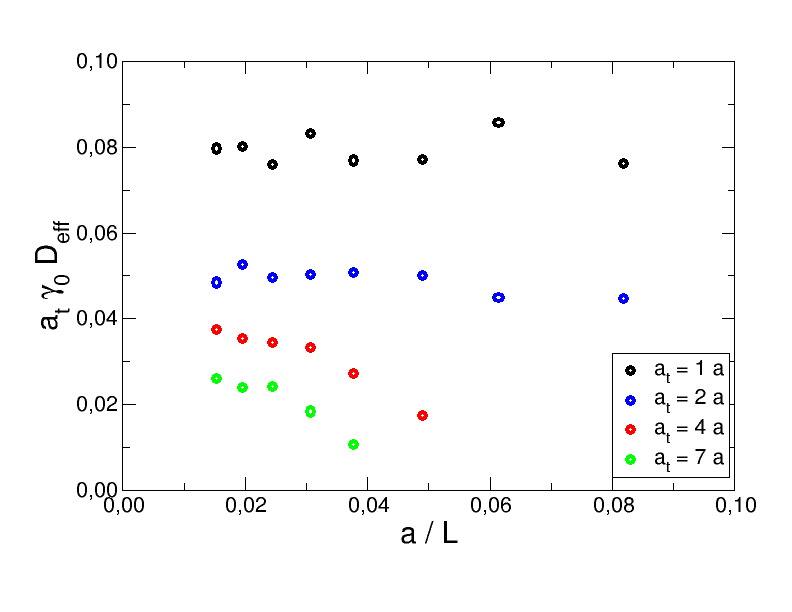,width=0.95\figurewidth}
\caption{Tracer self diffusion coefficient as function of the inverse simulation box size for different tracer sizes, as labelled.  \label{FSE-D}}
\end{figure}

Therefore, we first study the presence of FSE in the simulations. Fig. \ref{FSE-D} presents the tracer diffusion coefficient obtained in simulations  with different number of bath particles (in all cases, the volume fraction of the bath is $\phi=0.50$), and for different tracer sizes. The diffusion coefficient has been obtained in all cases from the long time slope of the tracer MSD. As mentioned above, the tracer volume is not accounted for; this provokes that $1/L$ is identical for all systems with the same $N$, but also that the minimum box size, maximal $1/L$, is larger for bigger tracers, as the system (tracer and bath particles) does not fit in the simulation box.

The results show that there are important FSE, but these do not follow the predictions mentioned above. In fact, for $a_t=a$ there are no FSE, whereas for large tracers, they appear but do not show the expected linear dependence on the inverse box size. For large simulation boxes, the FSE show a weaker dependence on $1/L$, or saturate. Similar results were obtained in active microrheology with a small force \cite{Orts2020}, and the effect was attributed to the Langevin microscopic dynamics, where particle momentum relaxes due to the solvent, breaking the $1/L$ dependence for Newtonian fluids. Thus we take the bath with $N=8000$ particles as representative of the "infinite" bath, and study the dynamics of the unforced tracer.

\subsubsection{Tracer motion}

\begin{figure}
\psfig{file=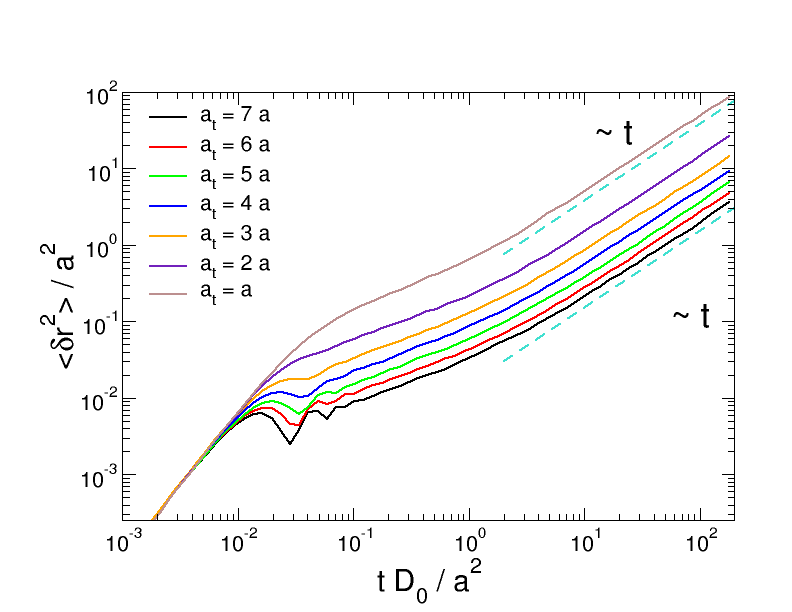,width=0.95\figurewidth}
\caption{Tracer MSD for different tracer sizes, as labelled, (increasing from the upper curve to the lower one) for the bath with $N=8000$ particles. \label{MSD-F0}}
\end{figure}

Fig. \ref{MSD-F0} shows the tracer MSD for tracer sizes from $a_t=a$ (equal to the bath particles) to $a_t=7a$. This shows that diffusion is attained at long times in all cases, after the collisions with the bath particles and friction with the solvent. For large tracers, the collisions provoke a rattling at intermediate times, before long time diffusion is reached, as noticed by the oscillations in the MSD. The diffusion coefficient (compared below with the friction coefficient from active microrheology with low forces) decreases with increasing the tracer size. These observations are confirmed by the theory, as shown below.

\begin{figure}
\psfig{file=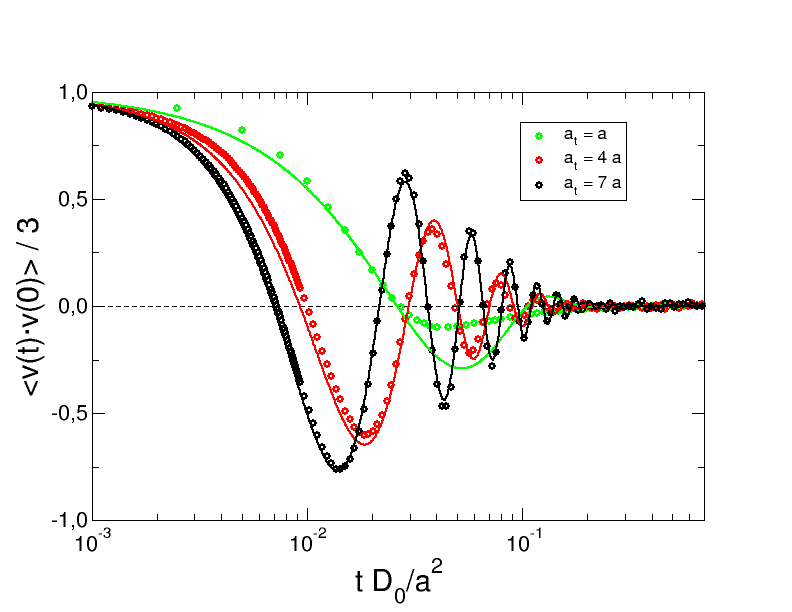,width=0.9\figurewidth}
\psfig{file=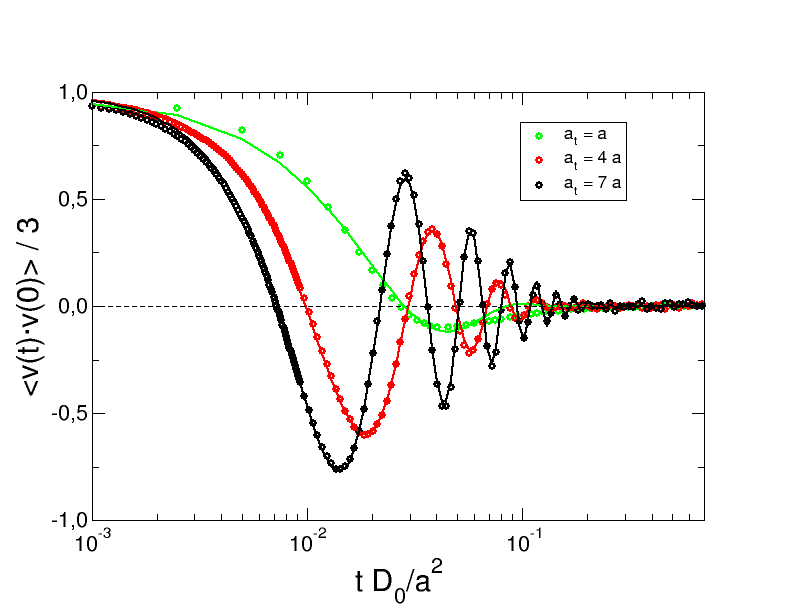,width=0.95\figurewidth}
\caption{Tracer VACF for different tracer sizes, as labelled, with $N=8000$ particles. The upper panel shows the fitting with the MCT model, and lower one with the damped harmonic oscillator, eq (\ref{harmonic-oscillator}). \label{vcorr}}
\end{figure}

The rattling of the tracer is also noticed by the tracer VACF, as shown in Fig. \ref{vcorr} for different tracer sizes. {The upper panel compares the simulation results with MCT, fitting $\Gamma$ to the damping as mentioned above; a further fitting of $D_0$ is needed to improve the match of theory and simulations. } The VACF for $a_t=a$ shows the typical shape of dense systems, with a fast decay to negative values, followed by a slow increase to zero. For larger tracer sizes, however, this transforms into an oscillatory curve with decreasing amplitude. Upon increasing the tracer size, the initial decay occurs at shorter times, the strength of the oscillations increases, and the global damping of the oscillations is weaker. {The MCT results overestimate the minimum of the $a_t=a$ case, but agree with the shape of the simulated VACF for larger tracers, describing the multiple oscillations accurately for large $a_t$.}

\begin{figure}
\psfig{file=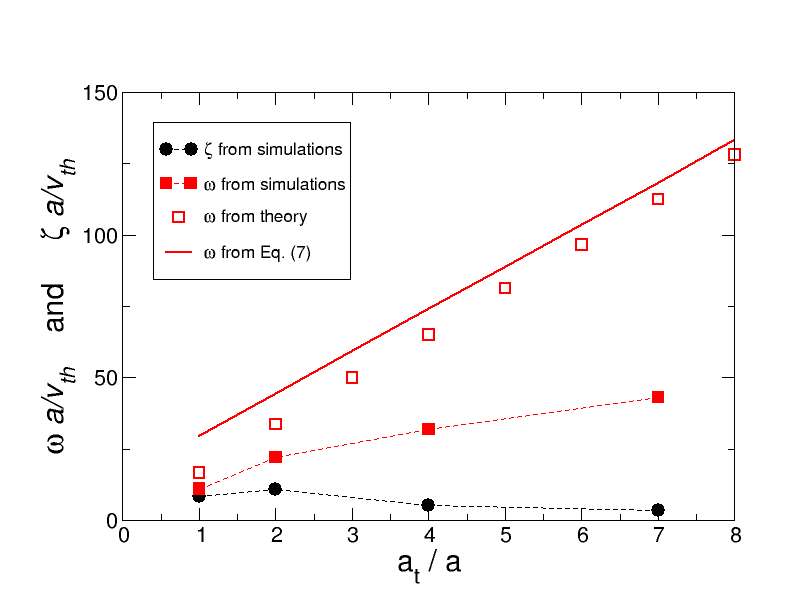,width=0.95\figurewidth}
\caption{Frequency and damping of the VACF from simulations and theory, as indicated.  \label{vcorr-params}}
\end{figure}

In order to characterize the VACF, and in particular the frequency of its oscillations, the overall shape of the VACF is described by a simple damped harmonic oscillator, as shown by the continuous lines in the lower panel of Fig. \ref{vcorr}, i.e. 

\begin{equation}
    \langle {\bf v}(\tau) \cdot {\bf v}(0) \rangle \; / 3  =  e^{-\zeta \tau} \cos \left( \omega \tau \right) \label{harmonic-oscillator}\;,
\end{equation}

\noindent with $\zeta$ and $\omega$ the friction and angular frequency of the damped oscillator, respectively. This expression can be fitted for all tracer sizes (both for the simulations and theory), and the evolution of the parameters $\gamma$ and $\omega$ with $a_t$ is shown in Fig. \ref{vcorr-params}. Note that the friction $\zeta$ is different from the friction with the solvent or the bath, as the decay of the VACF originates mainly from the momentum transport, although it is encoded here as a damping term. In the simulations, $\omega$ increases almost linearly for $a_t\geq 2a$ whereas $\zeta$ has a maximum for $a_t=2a$ and then decreases for increasing tracer sizes. 

A similar analysis of the numerical solutions of MCT, Eq.(\ref{MCT-K}), with fits of similar quality, has been performed (not shown). This provides the open symbols in Fig.~\ref{vcorr-params}, while the MCT solution for asymptotically large tracers, Eq. (\ref{MCT-w}), is shown as the thick continuous line, with good agreement between both sets of data in the large tracer limit. The discrepancy in $\omega$  between theory and simulation is corrected by adjusting the overall time scale, viz.~$D_0$, in the theory in the upper panel of Fig. \ref{vcorr}, as mentioned above. MCT thus points to the local structure encoded in the partial structure factors as origin of the tracer vibrations. They arise from the cage effect that the tracer gets scattered back from the shell of neighboring bath particles. {The differences between the simulation and theory data, which is accounted for in Fig. \ref{vcorr} via a time scaling factor, can be attributed to the approximations in the model such as considering a one-component system, using the PY structure factor, { neglecting the velocity fields of the bath particles around the tracer,}  and using the cage-effect description developed for capturing the glass transition at high packing fraction.}


\begin{figure}
\psfig{file=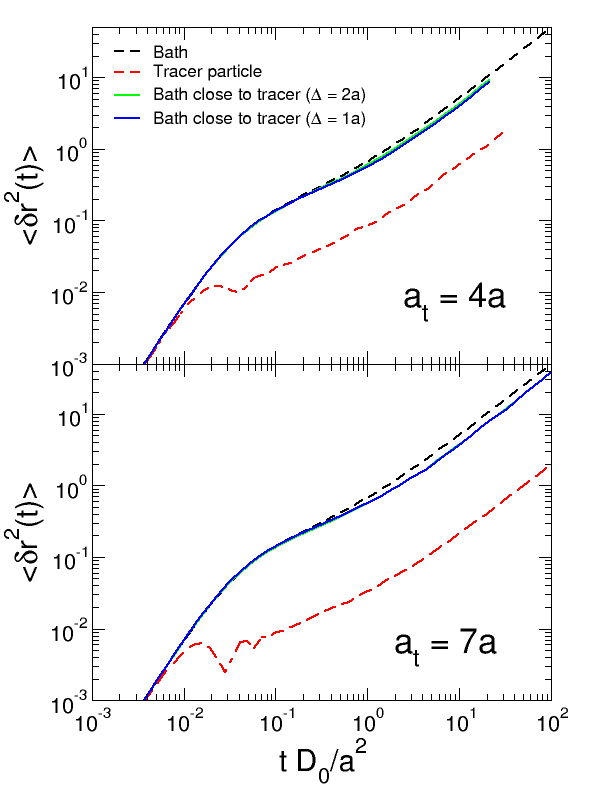,width=0.95\figurewidth}
\caption{Mean squared displacements of the bath particles without tracer (black dashed line), of the bath particles close to the tracer (red and green lines, as labelled), and of the tracer (dashed red line). The upper panel shows the results for $a_t=4a$ and the bottom one for $a_t=7a$. \label{msd-bath}}
\end{figure}

\subsubsection{Effects on the bath}

A critical approximation in the theory is that the dynamics of the bath particles is unaffected by the tracer. This is tested in this section by analysing the dynamics of bath particles close to the tracer.
 Fig. \ref{msd-bath} studies the MSD of these particles at a distance shorter than $\Delta$, for $a_t=4a$ and $7a$ (upper and lower panel of the figure, respectively), without external force. The figure also shows the MSD of the tracer and the bath particles in bulk (without the tracer). The comparison shows that even for low values of $\Delta$ ($\Delta=1a$), the dynamics of the bath particles resembles closely the bath without the tracer, and is notably different from the tracer dynamics. This indicates that the bath dynamics is mostly unaffected by the presence of the unforced tracer, thus confirming the approximation in MCT.

\subsection{Active microrheology. Linear and non-linear response}

\subsubsection{Finite size analysis} 

\begin{figure}
\psfig{file=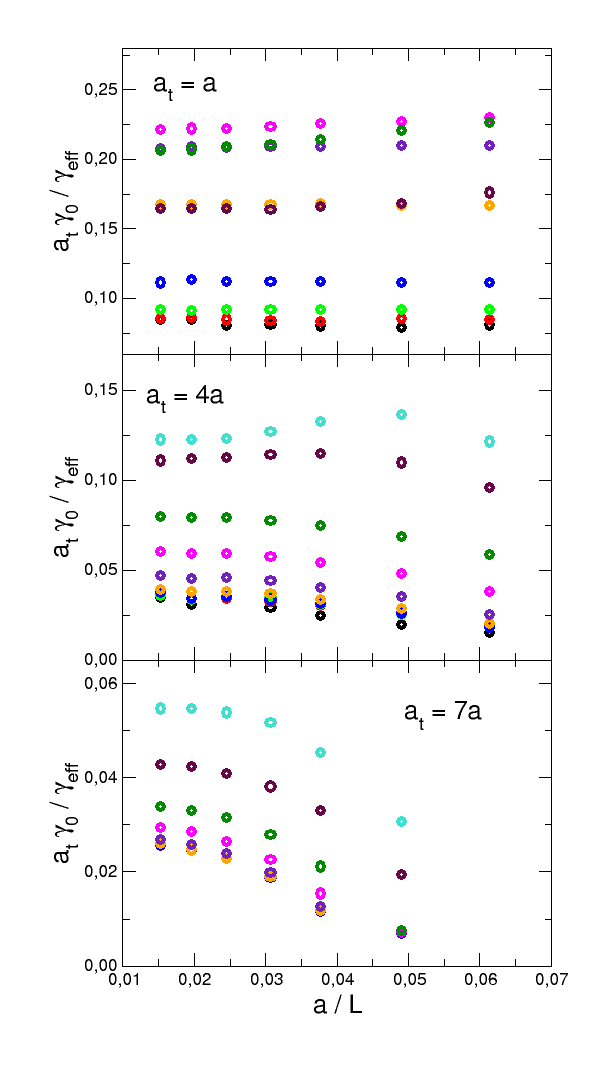,width=0.95\figurewidth}
\caption{Inverse friction coefficient as function of the inverse simulation box size for different forces, as labelled (increasing from bottom to top). The upper, intermediate and lower panels show $a_t= 1 a, 4a$ and $7a$, respectively.\label{gammaL}}
\end{figure}

We study now the case of the forced tracer, namely, active microrheology, starting again with the finite size effects. Fig. \ref{gammaL} shows the friction coefficient determined from the steady tracer velocity at long times for different forces and system sizes; every panel shows a value of the tracer radius. For all system sizes and tracer radii, the effective friction decreases upon increasing the pulling force. Also, in agreement with Fig. \ref{FSE-D}, the case of $a_t= 1 a$ (upper panel) has negligible FSE, while these are much more important for increasing tracer sizes. And again, these FSE are different from the theoretical expectation of a tracer in a Newtonian solvent \cite{Hasimoto1959}, which should be linear in this representation. As in the case of passive microrheology, the data apparently saturates for increasing number of particles.

\begin{figure}
\psfig{file=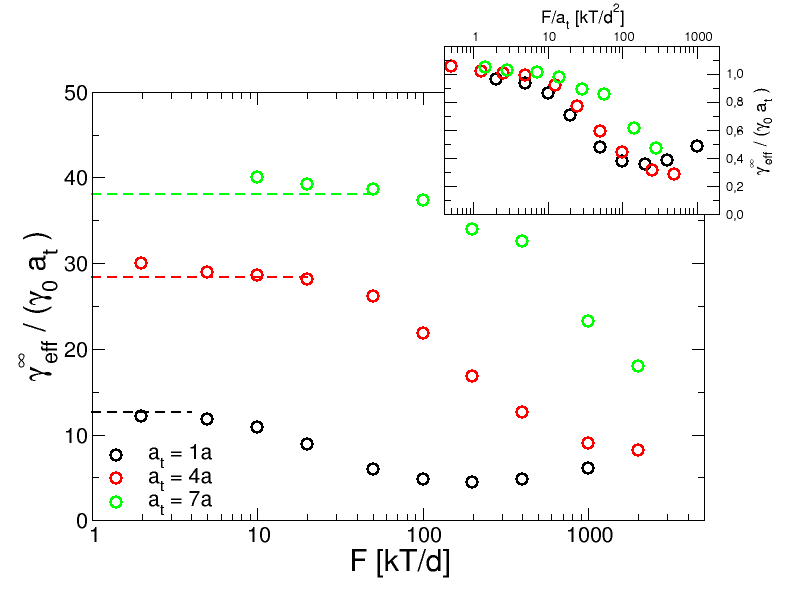,width=0.95\figurewidth}
\caption{Effective friction coefficient as function of the external force for different tracer sizes, as labelled (increasing from bottom to top).\label{gammaF}}
\end{figure}

The values of the friction coefficient in the biggest system are therefore considered as  "infinite-system" values, and are presented in Fig. \ref{gammaF} as a function of the external force for three tracer sizes. In all cases, the friction shows a linear regime at small forces, which grows and extends to stronger forces for larger tracers. The dashed horizontal lines at small forces represent the prediction from passive microrheology, obtained from the self diffusion coefficient, which agrees with the friction coefficient at low forces.

Upon increasing the force pulling the tracer, a force-thinning regime is found, where the tracer moves easier; the thinning amplitude increasing with the tracer size. Finally, the thinning terminates in a minimum followed by a slight increase (noticeable only for $a_t=a$). This stage has been described previously by Sperl et al. \cite{Sperl2016}, and can be attributed to collisions between the tracer and bath particles in the ballistic regime. 

The inset to the figure shows the same data, rescaling the effective friction coefficient with the self diffusion coefficient, and the force with the tracer size. This simple scaling accounts almost for the differences in the data, but the collapse onto a master curve, particularly of the thinning regime, is not fully attained.

\begin{figure}
\psfig{file=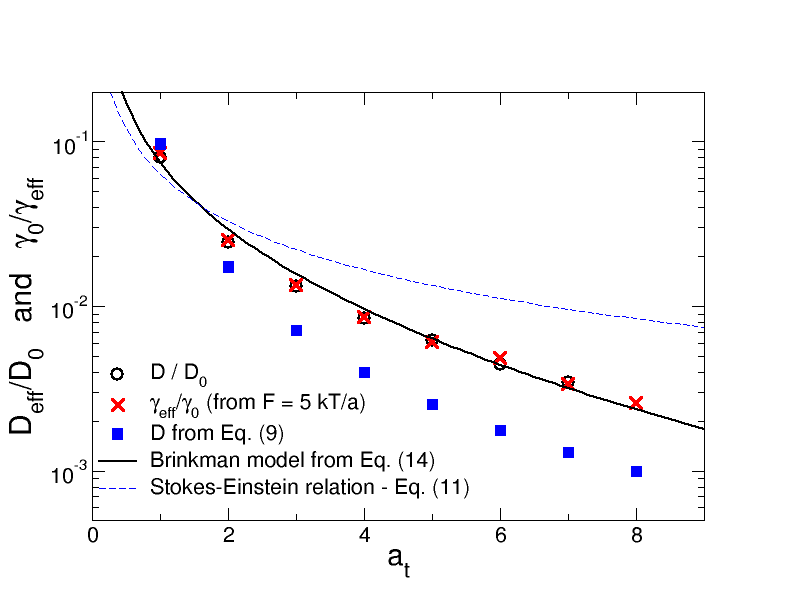,width=0.95\figurewidth}
\psfig{file=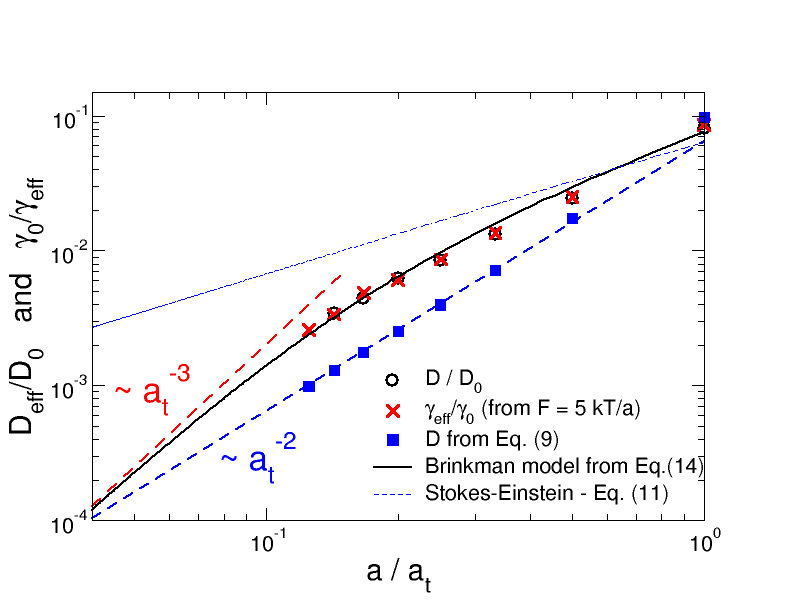,width=0.95\figurewidth}
\caption{Self diffusion coefficient and inverse effective friction coefficient for a small force, as function of the tracer sizes, as labelled. The results for the diffusion coefficient from the theory are also included as blue squares. 
The lower panel plots the same data as a function of $a/a_t$, and the power laws signalling the large tracer limit of the Brinkman model and MCT. 
\label{Dgamma}}
\end{figure}

\subsubsection{Force-motion relations} 

The comparison between passive and active microrheology (at low forces) is further studied in Fig. \ref{Dgamma}, by comparing the self diffusion coefficients and inverse friction coefficients. Both data sets agree perfectly, as anticipated by the data in Fig. \ref{gammaF}.
  The thin dashed blue line shows the prediction from the Stokes-Einstein relation,
\begin{equation}
   D = \frac{kT}{6 \pi \eta a_t+\gamma_0}\;, \label{SES}
\end{equation}

\noindent where $\eta$ is the shear viscosity of the bath, calculated from the Green-Kubo relation with the stress autocorrelation function. To reduce the numerical errors, the viscosity is calculated using the Einstein relation with the non-diagonal terms of the stress tensor \cite{Allen1987,Puertas2007}, yielding $\eta=(3.9 \pm 0.1) \sqrt{kT\,m}/a^2$. The prediction (thin dashed blue line), however, deviates clearly from the data, showing that the system under study does not correspond to a particle in a Newtonian solvent. 

In fact, a tracer in a bath of Brownian particles has been recently described using the following hydrodynamic equation \cite{Orts2020}, as proposed by Vogel et al. \cite{Vogel2019}:

\begin{equation}\label{brinkman}
\nabla P - \tilde{\eta} \nabla^2 {\bf u} = \tilde{\gamma}_0 {\bf u} + {\bf F}_{ext}\;,
\end{equation}

\noindent where $P$ is the pressure, ${\bf u}$ is the bath particles velocity field, and ${\bf F}_{ext}$ is the force pulling the tracer. This equation is formally similar to the Brinkman model \cite{Brinkman1947}, although the interpretation is different \cite{Orts2020}; the term with the Laplace operator describes the bath, and $\tilde{\gamma}_0 {\bf u}$ corresponds to the solvent. Therefore, we identify $\tilde{\eta} = \eta$, the viscosity of the collodial bath and  $\tilde{\gamma}_0 = n\gamma_0$, the friction coefficient with the solvent. The solution of this equation gives a linear dependence between the force ${\bf F}_{ext}$ and tracer velocity, but the exact expression depends on the boundary conditions on the tracer surface. For stick boundary conditions, this is given by \cite{Pozrikidis2011}:

\begin{equation}
{\bf F}_{ext} = 6 \pi \eta a_t {\bf u}_0 \left( 1 + k_0 a_t + \frac{1}{9} k_0^2 a_t^2 \right) ,
\end{equation}

\noindent with $k_0 = \sqrt{\tilde{\gamma}_0/\tilde{\eta}}$. On the other hand, if slip boundary conditions are assumed, the relation is:

\begin{equation}
{\bf F}_{ext} = 6 \pi \eta a_t {\bf u}_0 \left( \frac{2 + 2 k_0 a_t}{3 + k_0 a_t} + \frac{1}{9} k_0^2 a_t^2 \right).
\end{equation}

Defining the effective diffusion coefficient via $u_0 = \frac{D_{eff}}{kT} F_{ext}$, the results for best fitting is included in Fig. \ref{Dgamma}, using $k_0$ as fitting parameter. This is obtained for the slip boundary conditions, with a value of $k_0 = 0.55\,a^{-1}$, in semi-quantitative agreement with the expectation $k_0=\sqrt{n\gamma_0/\eta} = 0.39\,a^{-1}$, with $n$ the number density of bath particles (the fitting with the stick boundary conditions yields $k_0=0.078\,a^{-1}$, deviating more from the expectation).

Fig.~\ref{Dgamma} includes also the diffusion coefficients from the MCT of the  the glass transition in Brownian mixtures (Eq.~\ref{eq5}), as discussed above. These results correspond to the unforced tracer, and have been obtained from the long-time slope of the MSD. The diffusion coefficient follows the trend of both $D_{eff}$ and $1/\gamma_{eff}$, confirming the decrease with the tracer size. However, there are quantitative differences in the figure that show an overestimation of the slowing down of the tracer dynamics in the theory.

The lower panel of Fig. \ref{Dgamma} shows the same data as a function of $a/a_t$ to highlight the large-tracer limit. While the behaviour of the Brinkman model is $D_{eff} \sim a_t^{-3}$, Eq.~\eqref{eq5} predicts $D_{eff} \sim a_t^{-2}$ (both power laws are shown in the panel). The simulation data agrees  more closely with the Brinkman model, although the $a_t^{-3}$ limit is not reached, confirming that the MCT of the glass transition does not include the correct mechanism leading to tracer diffusion at low packing fractions, namely, transversal flow, as described above.

\subsubsection{Forced tracer motion} 

The formalism of linear response theory (LRT) not only provides the stationary state, and in particular the diffusion coefficient, as tested so far. The time-dependent response can also be calculated within LRT \cite{Kubo1957}, such as the {transient} tracer velocity \cite{Hansen1986,Leitmann2018}:

\begin{equation} 
\langle v_t(t) \rangle \:= \: \frac{\beta}{3}  F_{ext} \int_0^t \langle {\bf v}(t)\cdot {\bf v}(0) \rangle_{eq}\;, \label{LRT}
\end{equation}

\begin{figure*}
\psfig{file=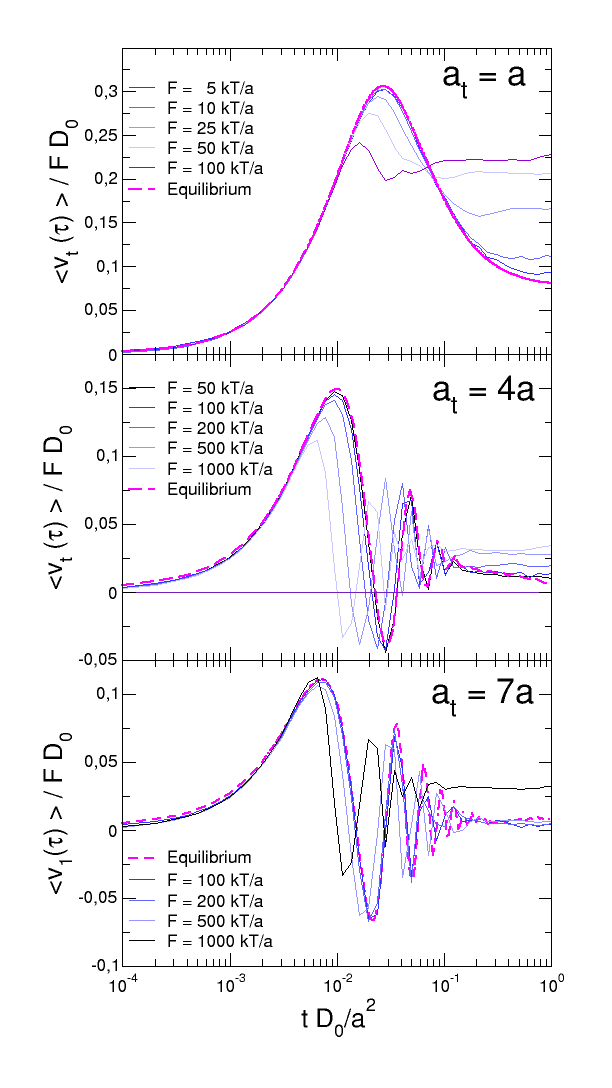,width=0.95\figurewidth}
\psfig{file=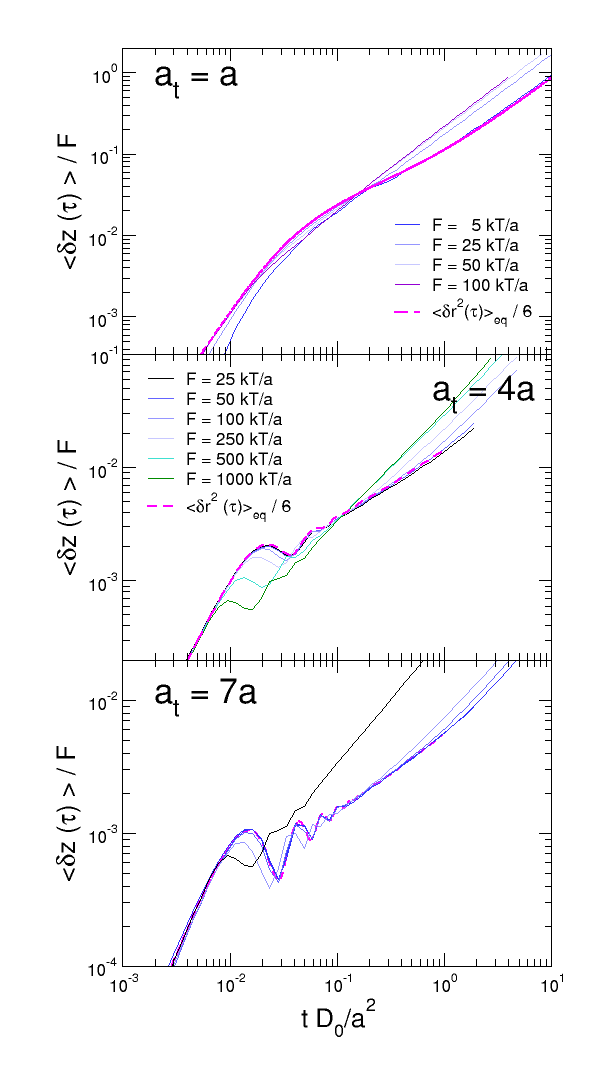,width=0.95\figurewidth}
\caption{Tracer velocity after the application of the external force for different forces, as labeled. The panels correspond to different tracer radius, $a_t = a$ (top panel), $a_t=4a$ (intermediate panel) and $a_t=7a$ (bottom panel). The dashed lines correspond to the integral in Eq. (\ref{LRT}). \label{vtracer}}
\end{figure*}

\noindent where it is assumed that the external force starts at $t=0$, i.e. $F(t) = F_{ext} \theta(t)$, with $\theta(t)$ is the Heaviside function, and $\langle {\bf v}(t) {\bf v}(0) \rangle_{eq}$ is the tracer VACF in equilibrium, i.e. without external force, shown in Fig. \ref{vcorr}. Alternatively, this relation can be integrated to yield:

\begin{equation} 
\langle z_t(t) \rangle \:= \: \frac{\beta F_{ext}}{6} \langle \delta r^2(t) \rangle_{eq} \;,\label{LRT-msd}
\end{equation}

\noindent with the same assumption for the external force. Both relationships, eqns. (\ref{LRT}) and (\ref{LRT-msd}), test the linear response approximation focusing on different aspects and therefore provide complementary information.

Fig. \ref{vtracer} tests both predictions for tracer radii $a_t=a$, $4a$ and $7a$, in the upper, intermediate and lower rows of panels, respectively. The left column of panels shows the tracer velocity (divided by the force) as a function of time after the start of the application of the force for different external forces, whereas the right column studies the tracer displacement and compares it with the equilibrium MSD. As visible in the figure, the tracer first accelerates in all cases, until it reaches the neighbouring bath particles, and after a transient regime, reaches a stationary velocity. However, whereas for $a_t=a$, the tracer velocity decreases after this initial acceleration, for larger sizes the velocity presents oscillations before the stationary state is reached. For small forces, the prediction of LRT is followed, either in the tracer displacement or in the velocity. This links  the oscillations in the velocity to the rebounds of the {forced} tracer in the cage of neighbours.

Upon increasing the external force, the initial acceleration of the tracer is unaltered, but the deviation from the equilibrium calculation is observed for shorter times and the oscillations have a shorter period. Note that the differences with respect to LRT occur both in the transient and stationary regimes, showing the transition from the linear to the non-linear regimes. In the transient velocity, this is noted as the failure of the LRT calculation, whereas in the stationary regime, it appears as the deviation from the low force plateau of the microviscosity, as shown in Fig. \ref{gammaF}. Finally, let us mention that for very large forces, the deviation from the equilibrium occurs at extremely short times, within the regime of ballistic motion, and causing the increase of the effective friction coefficient, in agreement with previous results \cite{Sperl2016}.

\subsubsection{Effects on the bath}

\begin{figure}
\psfig{file=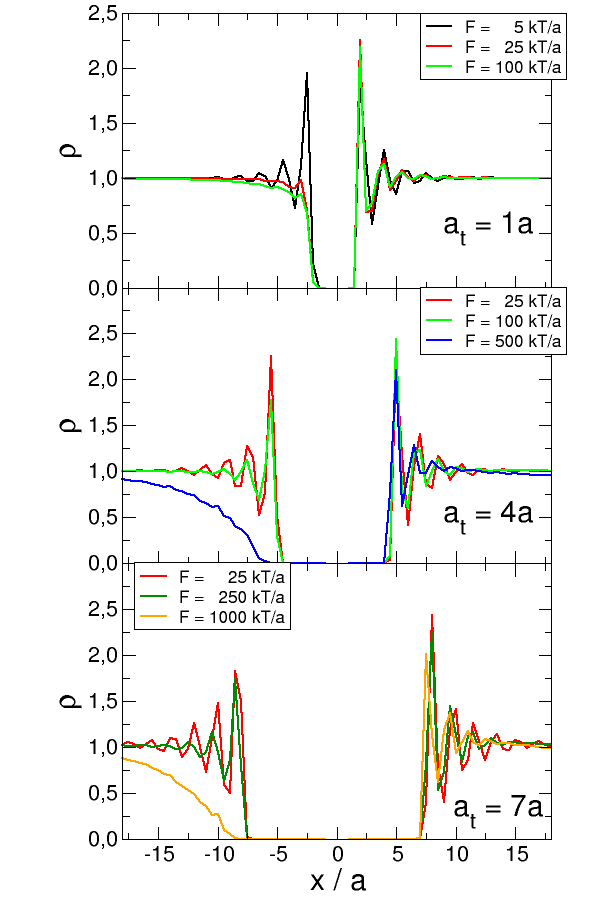,width=0.95\figurewidth}
\caption{Density of bath particles in the direction of the external force (in front of and behind the tracer), for different forces, as labeled. The panels correspond to different tracer radius, $a_t = a$ (top panel), $a_t=4a$ (intermediate panel) and $a_t=7a$ (bottom panel). \label{rho-bath}}
\end{figure}

In this subsection, the effects of the tracer motion on the bath are studied. Fig. \ref{rho-bath} presents the density of bath particles around the tracer, in the direction of the force (in front of and behind the tracer), in the stationary state (at long times) for the system with $N=8000$ particles. It must be recalled that no volume correction due to the tracer is considered, i.e. the bath is compressed. Tracer radii $a_t=a$, $4a$ and $7a$ are studied, for different forces in each case, covering the linear regime at low forces, the force thinning and the large force region. For small forces, the profile shows the typical oscillations due to the excluded-volume core of the bath particles. Also, it is symmetric around the tracer, because the perturbation induced by the moving tracer is almost unnoticed.

For larger forces, Fig. \ref{rho-bath}  shows that the most relevant effects in the bath appear downstream (behind the tracer). There, the oscillations are damped upon increasing the force, and then a wake appears when the tracer velocity is very large and the time scale of the moving tracer is much smaller than the time needed by the bath particles to fill the void behind the tracer. Upstream, minor changes in the density are also observed; the oscillations are compressed indicating the increased pressure. They are also damped at shorter lengthscales, due to the strain field induced by the tracer. The bath structure is quantitatively characterized by fitting the upstream density with:

\begin{equation}
\rho(x)\:=\: A e^{-\kappa (x-(a_t+a))} \cos \left[2\pi (x-(a_t+a))/ \lambda  \right] \;,\label{rho_bath-eq}
\end{equation}

\noindent which is motivated by the exponential decay predicted theoretically for the low density limit \cite{Squires2005}. 

\begin{figure}
\psfig{file=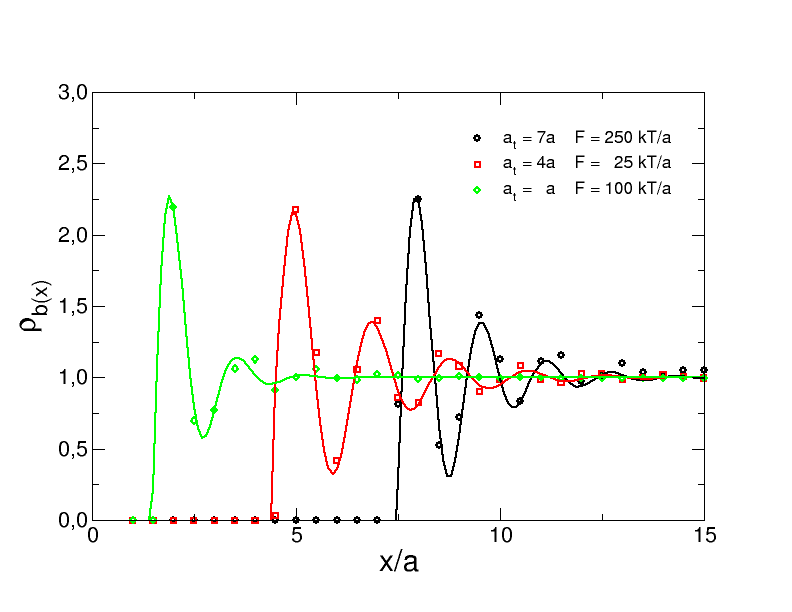,width=0.95\figurewidth}
\psfig{file=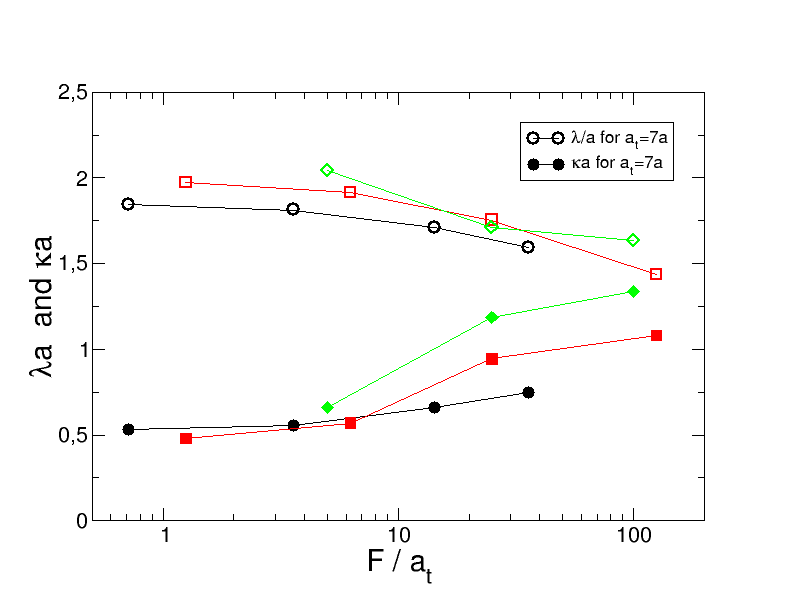,width=0.95\figurewidth}
\caption{Analysis of the density of bath particles in front of the tracer with size $a_t=7a$: Fitting of the density with expression \ref{rho_bath-eq} for different tracer sizes and forces, as labeled (upper panel) and dependence of the fitting parameters as a function of the external force, as labeled. \label{rho-fitting}}
\end{figure}

\begin{figure}
\psfig{file=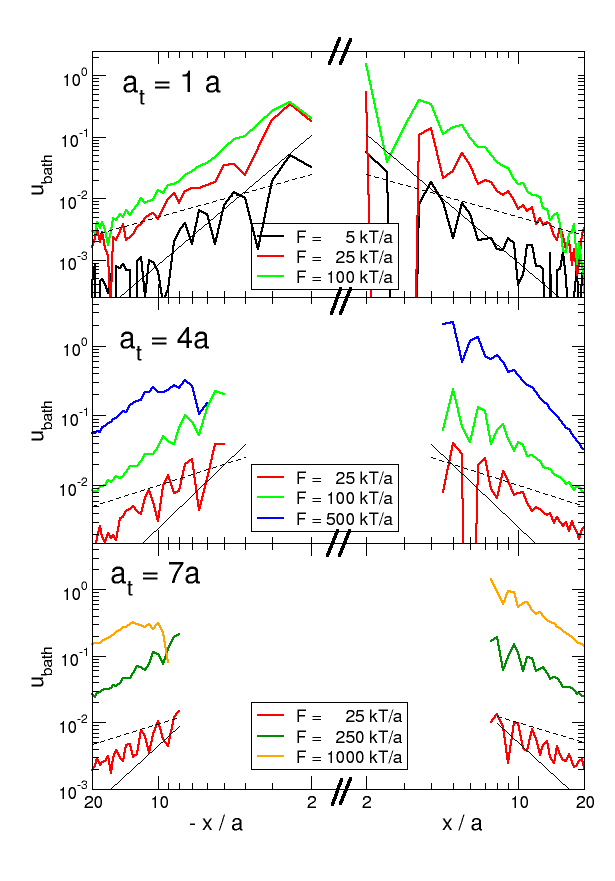,width=0.95\figurewidth}
\caption{Velocity of the bath particles behind and in front of the tracer for the same forces and tracer sizes as Fig. \ref{rho-bath}. The thin continuous black lines represent the $1/r^3$ behaviour, and the dashed ones the $1/r$ decay. \label{v-bath}}
\end{figure}

Fig. \ref{rho-fitting} shows the fitting of this expression to the bath density from the simulations for three typical cases in the upper panel. The analytical expression provides a good description of the data, and thus has been used to quantify the effects of the pushing tracer. The fitting parameters for all tracer sizes and forces are presented in the lower panel, as a function of $F/a_t$ to allow comparison between different tracers. Upon increasing the external force, the wavelength of the oscillations, $\lambda$, decreases slowly to values well below $2a$, whereas the damping parameter $\kappa$ increases continuously for all tracer sizes. As in the case of the friction coefficient (inset in Fig. \ref{gammaF}), plotting the data for different tracers as a function of $F/a_t$ nearly collapses the data onto a master curve, particularly in the case of $\lambda$.

Finally, we turn our attention to the dynamics of the bath particles. The motion of the particles close to a passive tracer remains little affected,  recall Fig.~\ref{msd-bath}. However, in active microrheology, the moving tracer modifies the dynamics of the bath. The velocity field in the bath is studied in Fig. \ref{v-bath} for the same cases as in Fig. \ref{rho-bath}, and compared with the theoretical predictions for the Brinkman and Navier-Stokes models (continuous and dashed black lines, respectively). The velocity profile decays far from the moving tracer, and increases with the external force both behind and in front of the tracer. More notably, the profile  oscillates in phase with the density, showing the difference between a continuum system and a particle based fluid with microscopic structure. 

Whereas the velocity field in the Brinkman model decays as $1/r^3$, in a Newtonian solvent (Navier-Stokes equation) it decreases as $1/r$. Both models are shown in the graphs for the lowest force and all tracer sizes (continuous and dashed lines, respectively). For small tracers, the velocity field from the simulations follows the prediction for the Brinkman model more closely, but differences are noticed for large tracers, although the $1/r$ decay is never followed. It is also interesting to note that for small forces the velocity profile is symmetric (upstream and downstream profiles are similar), whereas differences are observed for large forces. Notably, the velocity immediately behind the tracer is smaller than in front of it.  

The slower decay of the velocity profile for larger tracers provides an explanation for the stronger finite size effects for larger tracers, as presented in Figs. \ref{FSE-D} and \ref{gammaL}. For smaller tracers, the velocity profile decays at shorter distances, within the simulation box, whereas for large tracers, the slower decay provokes the effective interaction of the tracer with its periodic image, resulting in a higher friction coefficient, or smaller diffusion coefficient.

\section{Conclusions}

The dynamics of a spherical tracer in a bath of colloidal particles has been studied with Langevin dynamics simulations. Different sizes of the tracer have been considered, equal or larger than the bath particles in all cases. The analysis of finite size effects has shown an important influence of the number of bath particles in both active and passive microrheology, particularly for large tracers. However, these effects saturate for large enough systems, with $N=8000$ particles for a bath volume fraction of $\phi=0.50$. This holds because of Langevin dynamics and is in contrast with the predictions from the Navier-Stokes equation. Yet it allows the study of the tracer dynamics in large but finite size overdamped systems.

The results for the freely diffusing tracer in the bath (passive microrheology) show that diffusion is reached at long times in all cases. For large tracers, an intermediate regime where the tracer is transiently caged appears, with oscillations of the tracer MSD. 
Quantitative MCT calculations with the PY structure factor approximation show that the intermediate caging is the cause of the slowing down of the tracer motion. Caging is the hallmark of the structural relaxation in dense fluids. Here, it is also accompanied by vibrations in the tracer motion, which become stronger with increasing tracer size. For the studied tracer sizes, the size-dependence of the vibration frequency and diffusion coefficient are captured {qualitatively} in MCT, albeit with  errors  that increase with tracer size. {In the case of the vibration frequency, MCT predicts the correct asymptotic scaling with tracer size, but not for the diffusion coefficient.} The latter is well described by the Brinkman equation which is consistent with the shear viscosity kernel of MCT. Yet, a unified MCT of cage effect and transverse modes covering the density from dilute to glass transition is still missing.

In active microrheology, the phenomenology is even richer, with the system featuring linear and non-linear response. The friction coefficient shows a linear regime at small forces, followed by a decrease at intermediate ones, for all tracer sizes. In the large force limit, a shallow minimum is described, due to the interplay of inertial effects. LRT is used to calculate the tracer velocity and displacement upon the application of the external force with the VACF and MSD of the tracer in passive microrheology, respectively. The results for different tracer sizes are tested with the simulations of active microrheology, showing perfect agreement. In the stationary state, this yields the Stokes-Einstein relationship between the diffusion and friction coefficients (from passive and active microrheology, respectively), which is verified for all tracer sizes. 

The density of bath particles around the tracer shows the typical oscillations due to the particles' quasi-hard cores. When the tracer moves, it breaks the isotropy, and the oscillations in its front decrease in wavelength and are damped more strongly. This indicates that the bath is compressed in front of the moving tracer, and the local ordering is broken. Behind the tracer, a depletion zone or wake appears at very strong forces. The velocity profiles in the bath also present oscillations in phase with the density, but decay with the distance to the tracer faster than the prediction from Newtonian hydrodynamics, $1/r$. On the other hand, following the bath density, the velocity of the bath particles is smaller behind the tracer than in the front.

The friction coefficient and velocity profiles of the bath have been described using the Brinkman model. This was originally proposed to describe the dynamics of a tracer in a swarm of colloidal particles, and also used to study the diffusion in porous media. However, different from the original Brinkman model where the divergence of the stress tensor represents the solvent, here the solvent is described by a constant friction. This model reproduces the results of the Langevin dynamics simulations for small tracers or forces; in particular, the tracer diffusion coefficient is well fitted, and the decay of the velocity profile is correctly described. For large tracers, deviations from the behaviour of the Brinkman model are observed in the velocity profile, which decays more slowly, but faster than the hydrodynamic decay with the inverse distance, although a finite size effect cannot be ruled out.

\section{Acknowledgements}
AMP, GO and EMG acknowledge financial support thrrough projects PID2021-123278OB-I00 and PID2021-127836NBI00 (funded by MCIN/AEI/10.13039/501100011033/ FEDER “A way to make Europe”) and UAL2020-TIC-A2101 (funded by Junta de Andalucía and the European Regional Development Fund, ERDF). M.M. acknowledges support by the Deutsche Forschungsgemeinschaft (DFG, German Research Foundation) in project FU 309/12 and M.F. in SFB 1432, Project-ID 425217212, No. C06.

\bibliographystyle{apsrev}

\end{document}